\documentclass[prd,twocolumn,amsmath,amssymb,nofootinbib,preprintnumbers,balancelastpage]{revtex4}
\usepackage{graphicx}
\usepackage{hyperref}
\hypersetup{
    pdfnewwindow=true,      
    colorlinks=true,        
    linkcolor=black,        
    citecolor=blue,         
    filecolor=blue,         
    urlcolor=blue           
}
\usepackage{cancel}
\usepackage{amssymb}
\usepackage{textcomp}
\usepackage{amsmath}
\usepackage{bm}
\usepackage{times}
\usepackage{epsfig}
\usepackage{color}
\usepackage{graphics}
\usepackage{hyperref}
\usepackage{setspace}
\usepackage{slashed}
\usepackage{array}
\def\bea{\begin{eqnarray}}
\def\eea{\end{eqnarray}}
\begin{document}
\title{\Large {\it{\bf{Unification and Local Baryon Number}}}}
\author{Pavel Fileviez P\'erez$^{1}$, Sebastian Ohmer$^{2}$}
\affiliation{\vspace{0.15cm} \\  $^{1}$CERCA, Physics Department, Case Western Reserve University, Rockefeller Bldg. 2076 Adelbert Rd. Cleveland, OH 44106, USA
\vspace{0.15cm} \\  $^{2}$Particle and Astro-Particle Physics Division \\
Max-Planck-Institut fuer Kernphysik {\rm{(MPIK)}} \\
Saupfercheckweg 1, 69117 Heidelberg, Germany}
%
\date{\today}
\begin{abstract}
We investigate the possibility to find an ultraviolet completion of the simple extensions of the Standard Model where baryon number is 
a local symmetry. In the context of such theories one can understand the spontaneous breaking of baryon number at the low scale and the 
proton stability. We find a simple theory based on $SU(4)_C \otimes SU(3)_L \otimes SU(3)_R$ where baryon number is embedded in a 
non-Abelian gauge symmetry. We discuss the main features of the theory and the possible implications for experiments. This theory predicts 
stable colored and/or fractional electric charged fields which can give rise to very exotic signatures at the Large Hadron Collider experiments such as CMS and ATLAS.
We further discuss the embedding in a gauge theory based on $SU(4)_C \otimes SU(4)_L \otimes SU(4)_R$ which could define the way to 
achieve the unification of the gauge interactions at the low scale.   
\end{abstract}
\maketitle

\section{Introduction}
\label{intro}
The Standard Model (SM) of particle physics is today a very successful theory of nature which describes most of the experimental results below or at the electroweak scale, $\Lambda_{W} \sim 100$ GeV. There are many appealing ideas for physics beyond the Standard Model. In the context of grand unified theories one can understand why the weak, the electromagnetic 
and strong interactions are so different. In this context the SM interactions are just different manifestations of the same fundamental interaction~\cite{GG}. 
The idea of grand unification is very appealing but unfortunately these theories can be realized only at the high energy scale, $M_{GUT}\approx 10^{14-16}$ GeV, in agreement 
with evolution of the gauge couplings~\cite{GQW} and the proton decay experimental bounds~\cite{proton}. It is difficult to imagine a direct test of these theories at colliders 
due to the fact that they can only be realized at energy scales much larger than the center-of-mass energy of any future collider experiment.

Recently, several effective field theories have been proposed in order to understand the possible absolute stability of the proton~\cite{B1,B2,B3,B4,Breview}. 
In this context the global baryon number symmetry present in the SM is promoted to be a local gauge symmetry which is spontaneously broken at the low scale.
Since in this context baryon number is not broken in one unit these theories predict that the proton is stable. Therefore, these theories provide an ideal framework 
to investigate the unification of fundamental forces at scales much smaller than the standard grand unified scale. See Refs.~\cite{pheno1,pheno2,pheno3,pheno4} for different studies in the context of these theories. In Ref.~\cite{pheno2} we have investigated the unification of gauge interactions from the bottom-up approach in order 
to understand the different scenarios where the unification can be realized at the low scale in agreement with the experiment.

In this article we investigate simple unified theories which are the ultraviolet completion of the theories proposed in Refs.~\cite{B1,B2,B3,B4} and can be realized at the low scale. 
These theories are crucial to understand the possibility to define a grand unified theory where the baryon number is a local symmetry and the proton is stable. In this way one can hope 
to have a realistic theory for the unification of gauge interactions at the low scale.
We focus mainly on a simple theory based on $SU(4)_C \otimes SU(3)_L \otimes SU(3)_R$ where baryon number is embedded in a 
non-Abelian gauge symmetry. In this context the color and the baryon number are unified in the $SU(4)_C$ symmetry.
We discuss the main features of the theory and the possible implications for experiments. This theory predicts 
stable colored and/or fractional electric charged fields which can give rise to very exotic signatures at the Large Hadron Collider experiments such as CMS and ATLAS.
We also discuss the embedding in a gauge theory based on $SU(4)_C \otimes SU(4)_L \otimes SU(4)_R$ which could define the way to 
achieve the unification of the gauge interactions at the low scale since the proton is absolutely stable in this context.  
%
\section{Theories for Baryon Number}
%
The idea of investigating a theory where the baryon number is defined as a local symmetry was discussed in 
Refs.~\cite{Lee:1955vk,Pais:1973mi}. See also Refs.~\cite{Rajpoot:1987yg,Foot:1989ts,Carone:1995pu} 
for previous studies. 
In Refs.~\cite{B1,B2,B3,B4} we have proposed several theories where the gauge symmetry is the SM gauge symmetry 
and a local Abelian symmetry $U(1)_B$. These theories have the common features:

\begin{itemize}

\item The local baryon number can be spontaneously broken at the low scale.

\item The proton can be absolute stable in the simplest models.

\item In order to define an anomaly free theory one needs to introduce extra vector-like fermions with different baryon numbers.

\item In this context one predicts the existence of a leptophobic gauge boson, $Z_B$, which couples only to the SM quarks and the new vector-like fermions.

\item These theories predict generically a cold dark matter candidate which is the lightest neutral field in the new sector of the theory. 

\item One typically has a relation between the matter-antimatter and the dark matter asymmetries.

\end{itemize}  
Clearly, these effective theories can provide an ideal framework to understand the unification of interactions at the low scale. 
In this article we propose a new theory which can help us to understand this possibility.  

\section{Towards Unification}
In the previous section, we have discussed the theories where baryon number is a local symmetry. 
Typically, in an unified theory where quarks and leptons live in the same multiplet one cannot 
define B and L for the matter multiplets. However, if one has extra matter one can have matter 
multiplets with definite baryon number and one of the generators in the algebra can be identified 
as baryon number generator. One example has been discussed in Ref.~\cite{Fornal:2015boa} 
where following the old idea of Pati-Salam~\cite{Pati:1974yy} and the model proposed in Ref.~\cite{Perez:2013osa}, 
$U(1)_B$ lives inside $SU(4)_C$. This idea is not new but it is the key to find a new simple 
unified theory where the baryon number is a local symmetry. The model proposed in Ref.~\cite{Fornal:2015boa} is simple but one cannot realize the breaking of the gauge symmetry 
to the SM model plus the local gauge symmetry $U(1)_B$ and the proton is not stable. One also has an Abelian gauge symmetry as in the SM. 

In this section, we show that it is possible to use the idea of  trinification to motivate a simple theory where baryon number is a local symmetry.  
We therefore consider the gauge group
\begin{equation}
G_{433}=SU(4)_C \otimes SU(3)_L \otimes SU(3)_R \,,
\end{equation}
with the Standard Model fermions embedded in
\begin{equation}
Q \sim (4, \bar{3}, 1) \,, \quad Q^c \sim (\bar{4}, 1, 3) \,, \text{and} \quad L \sim (1, 3, \bar{3})  \,.
\end{equation}
Anomaly cancellation requires that we introduce the extra fields
\begin{equation}
\Psi^c \sim (1, 3, 1) \quad  \text{and} \quad  \eta \sim (1, 1, \bar{3}) \,.
\end{equation}
Let us discuss the explicit form of all matter multiplets:
\begin{itemize}
\item The left-handed SM quarks live in the $Q$ multiplet (one for each family) together with extra colorless fields with baryon number as we show here 
\begin{equation}
Q = {\begin{pmatrix} d_r && u_r && D_r \\ d_b && u_b && D_b \\ d_g && u_g && D_g \\ \Psi_d && \Psi_u && \Psi_D \end{pmatrix}} =
{\begin{pmatrix} q \\  \Psi \end{pmatrix}}.
\end{equation}
Here $q$ is the quark multiplet used in trinification, which is the theory based on $SU(3)_C \otimes SU(3)_L \otimes SU(3)_R$ 
and $\Psi$ is a vector containing the extra fields $\Psi_d$, $\Psi_u$ and $\Psi_D$. 
The indices $r$, $b$ and $g$ correspond to the different colors.
\item The right-handed SM quarks live in the $Q^c$ multiplet together with the partners of the extra fields in $Q$
\begin{equation}
Q^c = {\begin{pmatrix} d^c_{\bar{r}} && d^c_{\bar{b}} && d^c_{\bar{g}} && \eta_d^c \\ u^c_{\bar{r}} && u^c_{\bar{b}} && u^c_{\bar{g}} && \eta_u^c\\ D^c_{\bar{r}} && D^c_{\bar{b}} && D^c_{\bar{g}} && \eta_D^c \end{pmatrix}} = {\begin{pmatrix} q^c && \eta^c \end{pmatrix}},
\end{equation}
\item The SM leptons live in the $L$ multiplet together with extra heavy leptons as in the case of trinification
\begin{equation}
L = {\begin{pmatrix}
N_1 && E^+ && \nu \\
E^- && N_2 && e^- \\
\nu^c && e^+ && N_3
\end{pmatrix}},
\end{equation}
\item As we have mentioned, here one needs the extra fermions $\Psi^c$ and $\eta$ to cancel all the anomalies. They are given by
\begin{equation}
\Psi^c = {\begin{pmatrix} \Psi^c_{d} && \Psi^c_{u} && \Psi^c_{D}  \end{pmatrix}} \quad \text{and} \quad \eta ={\begin{pmatrix} \eta_{d} \\ \eta_{u} \\ \eta_{D}  \end{pmatrix}}. 
\end{equation}
\end{itemize}
This theory has many interesting features:

\begin{itemize}

\item The decay of the proton could happen if the dimension nine operators, such as $QQQL\Phi\Phi\phi^\dagger/\Lambda^5$, are present. Therefore, the scale 
$\Lambda$ can be small. Here $\Phi \sim (1,3,\bar{3})$ is the scalar field present in trinification. See the next sections for details.

\item In the limit $v_B \to \infty$ the new fermions decouple and we are left with the known trinification model based on the $SU(3)_C \otimes SU(3)_L \otimes SU(3)_R$ gauge symmetry.

\item In the leptonic multiplet $L$ the leptons and anti-leptons are unified in the same representation. Therefore, the total lepton number is broken explicitly in this theory.

\item This theory predicts the existence of new vector-like quarks, $D + D^c$, with electric charge $-1/3$, new extra 
neutral leptons $N_1$, $N_2$, $N_3$ and the right-handed neutrinos $\nu^c$. We also find new vector-like heavy leptons $E^+$ and $E^-$ as in trinification.

\item There are exotic fields with fractional charge as well, the fields $\Psi$, $\Psi^c$, $\eta$ and $\eta^c$. The electric charges are given by
\begin{eqnarray*}
&Q(\Psi_u) = +2/3\,, Q(\Psi_d)=Q(\Psi_D) = -1/3, \\
&Q(\Psi^c_u) = -2/3\,, Q(\Psi^c_d)=Q(\Psi^c_D) = +1/3, \\
&Q(\eta_u) = +2/3\,, Q(\eta_d)=Q(\eta_D) = -1/3, \\
&Q(\eta^c_u) = -2/3\,, Q(\eta^c_d)=Q(\eta^c_D) = +1/3\,.
\end{eqnarray*}

\end{itemize} 
%
\subsection{Symmetry Breaking}
%
The gauge symmetry $SU(4)_C \otimes SU(3)_L \otimes SU(3)_R$ can be broken in two steps:
\begin{itemize}

\item $SU(4)_C$ breaks down to $SU(3)_C \otimes U(1)_B$ once the Higgs $\Sigma \sim (15, 1, 1)$ acquires a vacuum expectation value in the $T^{15}_C$ direction, 
$\langle \Sigma \rangle = v_C\, T^{15}_C$,  here $T^{15}_C=\text{diag}(1,1,1,-3)/2\sqrt{6}$. 
Therefore, one obtains a low energy theory based on trinification and the local baryon number, i.e. $SU(3)_C \otimes SU(3)_L \otimes SU(3)_R \otimes U(1)_B$. 
The Baryon number generator will be
\begin{equation}
Y_B = 2 \sqrt{\frac{2}{3}} T^{15}_C \,.
\end{equation}

\item The $U(1)_B$ gauge boson must acquire mass. We achieve it adding a new Higgs, $\phi \sim (4, 1, 1)$ which vacuum expectation value breaks the local baryon number.
The explicit form of the $\phi$ field is given by $ \phi^T = ( \phi_r  \  \phi_b \ \phi_g \ S_B )$ and the VEV of $S_B$ is $v_B/\sqrt{2}$ in our notation.

\end{itemize} 
We obtain a trinification theory based on $SU(3)_C \otimes SU(3)_L \otimes SU(3)_R$ which has the SM matter content plus extra matter fields and leptophobic gauge bosons. 
We will discuss in detail all the features of this model in the next sections.

The extra fields beyond trinification are $\Psi, \Psi^c, \eta$ and $\eta^c$, they acquire mass once $S_B$ gets a VEV. The relevant terms are  
\begin{equation}
\mathcal{L} \supset  - \left( y_\Psi \phi^\dagger Q \Psi^c \ + \ y_\eta \phi Q^c \eta + \text{h.c.} \right),
\end{equation}
and their masses read as
\begin{equation}
M_\Psi = y_\Psi \frac{v_B^*}{\sqrt{2}}, \  \text{and} \ M_\eta = y_\eta \frac{v_B}{\sqrt{2}}.
\end{equation}
The mass of the rest of the fermionic fields is protected by the trinification symmetry. We will discuss 
the effective trinification theory in the next sections. It is important to emphasize that both scales $v_C$ and 
$v_B$ can be in the multi-TeV region in agreement with all experimental constrains.
%
\subsection{Leptophobic Gauge Bosons}
%
This simple theory predicts the existence of seven extra leptophobic gauge bosons, $X^\mu$ and $Z_{B}^ {\mu}$, 
beyond trinification. In matrix form the $SU(4)_C$ gauge bosons can be decomposed as
\begin{equation}
A_\mu = {\begin{pmatrix} G_\mu && X_\mu /\sqrt{2} \\ X^*_\mu /\sqrt{2} && 0 \end{pmatrix}} + Z_{B\mu} T^{15}_C .
\end{equation}
The mass of the leptophobic gauge boson $Z_B$ is proportional to $v_B$, and the extra six gauge bosons acquire mass from the VEV of $\Sigma$. 
Breaking $SU(4)_C \to SU(3)_C \otimes U(1)_B$ we find the eight massless SM gluons $G_\mu \sim (8,1,1)$ which correspond to the first eight generators 
$T^{1\dots8}_C$, six massive gauge bosons corresponding to the generators $T^{9\dots14}_C$ which form three complex massive gauge bosons 
$X_\mu \sim(3,1,1)$ with masses at the symmetry breaking scale $v_B$ given by
\begin{equation}
M^2_X = \frac{2}{3} g_C^2 v_C^2 + \frac{1}{4} g_C^2 v_B^2 \,.
\end{equation}
Notice that in this model $g_B=\sqrt{3/8} \ g_C$ at the symmetry breaking scale $v_B$. These leptophobic gauge bosons can have the following decays $$X_\mu \to q \bar{\Psi}, \eta^c \overline{q^c}.$$ 
It is important to mention that the gauge fields $X_\mu$ can be light because they do not mediate flavour violating processes at tree level as the vector-leptoquark 
fields in models for quark-lepton unification, see for example Ref.~\cite{Perez:2013osa}. They will mediate flavour violation at one loop level where inside the loop you will 
have $X_\mu$ and the extra fermions, $\Psi$ or $\eta^c$. 

The other leptophobic gauge boson $Z_B$ will acquire a mass when $U(1)_B$ is spontaneously broken. The mass is given by
\begin{equation}
M_{Z_B} = g_B v_B,
\end{equation}
and they can decay as follows $$Z_B \to \bar{q} q, \bar{\Psi} \Psi, \overline{q^c} q^c, \overline{\eta^c} \eta^c.$$
These decays will be investigated in the next section.
%
\subsection{Cosmological Constrains}
%
Unfortunately, all the extra fields in this theory have color or fractional electric charge and then they can affect the predictions for cosmology.
Here we will assume that in order to avoid stable exotic fields with color or electric charge the reheating temperature is below the mass of the lightest 
exotic field. This problem is similar to the monopole problem in grand unified theories where after inflation we do not assume a reheating temperature 
of order GUT scale.

The theory predicts stable fractional charged fermions $\Psi$, $\Psi^c$, $\eta$ and $\eta^c$ or stable colored bosons $\phi_i$, with $i=r,b,g$ and $X^\mu$. 
We therefore have to require a reheating temperature below their masses to avoid the production of these fileds. There are three possible scenarios:
\begin{itemize}
\item Scenario A: One of the extra fields $\Psi$, $\Psi^c$, $\eta$ and $\eta^c$ is the lightest new field.
Then we have to require that the following decay channels are open in the early Universe
\begin{eqnarray*}
\phi_{i} & \to & \Psi + \Psi^c + X^\mu_i, \\ 
\phi_{i} & \to & \eta + \eta^c + X^\mu_i, \\
X^\mu_i &\to& q_i \overline{\Psi}, \overline{q_i^c} \eta^c \,.
\end{eqnarray*}
Therefore, in this case the lightest field has fractional electric charge and the upper bound on the reheating temperature is thus
\begin{equation}
T_{RH}^A << \frac{1}{\sqrt{2}} {y_{\Psi, \eta}} v_B \,.
\end{equation}

\item Scenario B: In this case the extra colored gauge boson, $X^\mu$, is the lightest new field.
Then, we have to require that the following decay channels are open in the early Universe
\begin{eqnarray*}
\phi_{i} &\to& \Psi + \Psi^c + X^\mu_i, \\
\phi_{i} &\to& \eta + \eta^c + X^\mu_i, \\
\Psi &\to& (X^\mu_i)^* \bar{q_i}, \\
\eta^c &\to& X^\mu_i \overline{q^c_i}.
\end{eqnarray*}
Then, the upper bound on the reheating temperature reads as
\begin{equation}
T_{RH}^B << g_C \sqrt{\frac{2}{3} v_C^2 + \frac{1}{4} v_B^2} \,.
\end{equation}
\item Scenario C: In this case $\phi_{i}$ is the lightest new field. 
The following channels have to be open in the early Universe
\begin{eqnarray*}
X^\mu &\to& \phi_{i} + S_B, \\
\Psi &\to& \bar{q} + \phi_{i}, \\ 
\eta &\to& q^c_i + \phi_{i} \,.
\end{eqnarray*}
Therefore, the upper bound on the reheating temperature reads as
\begin{equation}
T_{RH}^C <<  {M_\phi}.
\end{equation}
\end{itemize}
In all these scenarios the reheating temperature can be large, and we do not have any problems 
with BBN since it will be much larger than a few MeVs. We assume that the gauge symmetry is broken 
in the multi-TeV region and the leptophobic gauge boson can be produced at the LHC, this scenario 
is consistent with cosmology.
%
\section{Trinification at the low scale}
%
Once $SU(4)_C$ is broken to $SU(3)_C \otimes U(1)_B$ and $U(1)_B$ is broken,  
we find a simple trinification model which is based on $$SU(3)_C \otimes SU(3)_L \otimes SU(3)_R,$$ with extra matter. The Standard Model quarks and additional d-like quarks are embedded into $q$ and $q^c$ which are given by
\begin{equation}
q = {\begin{pmatrix} d_r && u_r && D_r \\ d_b && u_b && D_b \\ d_g && u_g && D_g \end{pmatrix}} \, \text{ and } \, q^c = {\begin{pmatrix} d^c_{\bar{r}} && d^c_{\bar{b}} && d^c_{\bar{g}} \\ u^c_{\bar{r}} && u^c_{\bar{b}} && u^c_{\bar{g}} \\ D^c_{\bar{r}} && D^c_{\bar{b}} && D^c_{\bar{g}}\end{pmatrix}}.
\end{equation}
The needed Higgses for symmetry breaking are
\begin{equation}
 \Phi_i = {\begin{pmatrix}
\varphi^0_{1i} && \varphi^+_i && H_{Li}^0 \\
\varphi^-_i && \varphi^0_{2i} && H^-_{i} \\
H_{Ri}^0 && H^+_i && S^0_i 
\end{pmatrix}} \sim (1,3,\bar{3})\,.
\end{equation}
The interactions of the quark fields and the scalar sector in trinification are described by
\begin{eqnarray}
- \mathcal{L} & \supset &\, q q^c (y_1 \Phi_1 + y_2 \Phi_2 + y_3 \Phi_3) + \nonumber \\
&& \Psi^c \eta (h_1 \Phi_1^\dagger + h_2 \Phi_2^\dagger + h_3 \Phi_3^\dagger )+ \text{h.c.} \,,
\end{eqnarray}
where the interaction between the quarks and one of the $\Phi$ scalar fields is given by
\begin{eqnarray}
q q^c \Phi & =& d d^c\varphi_1^0 + u u^c\varphi_2^0 + d u^c \varphi^+ + u d^c \varphi^- \nonumber \\
&-& \, dD^c H^0_L - Dd^c H_R^0 + Du^cH^+ + uD^c H^- + DD^c S^0. \nonumber \\
\end{eqnarray}
The masses of the SM quarks are generated once the $\varphi_1^0$ and $\varphi_2^0$ acquire a vacuum expectation value.
However, we need three copies of the $\Phi$ fields to generate a realistic spectrum for the SM quark masses, see 
Refs.~\cite{333-1,333-2,333-3,333-4,333-5,333-6} for more details.

The interactions of the leptonic fields with the scalar degrees of freedom are described by
\begin{equation}
- \mathcal{L} \supset \frac{1}{2} L L (k_1 \Phi_1 + k_2 \Phi_2 + k_3 \Phi_3 ) + \text{h.c.} \,,
\end{equation}
where $ L L \Phi = \epsilon^{ijk}\epsilon_{abc} L^i_a L^j_b \Phi^k_c$. Using these interactions one can generate SM 
lepton masses as we did for the SM quarks.
Once the $SU(3)_L \otimes SU(3)_R$ symmetry 
is broken, there are nine massive gauge bosons beyond the gauge bosons present in a left-right symmetric theory.
The extra massive gauge bosons have the following properties:
\begin{itemize}
\item Extra gauge bosons associated to $SU(3)_L$:
\begin{equation}
B_L^\mu \sim (1,\bar{2},1,-1)\quad \text{and} \quad \tilde{B}_L^\mu \sim (1,2,1,1)\,.
\end{equation}
\item Extra gauge bosons associated to $SU(3)_R$:
\begin{equation}
B_R^\mu \sim (1,1,\bar{2},-1) \quad \text{and} \quad \tilde{B}_R^\mu \sim (1,1,2,1) \,.
\end{equation}
\item Mixed massive gauge boson
\begin{equation}
C^\mu = c_{\theta_3} C_R^\mu - s_{\theta_3} C_L^\mu \sim (1,1,1,0)\,,
\end{equation}
\end{itemize}
where we use the abbreviations $s_{\theta_3}=\text{sin}\theta_3$ and $c_{\theta_3}=\text{cos}\theta_3$.
We find the following mass terms in the Lagrangian
\begin{equation}
- \mathcal{L} \supset \frac{M^2_{B_L}}{2} \tilde{B}_L^\mu B_{\mu L} + \frac{M^2_{B_R}}{2} \tilde{B}_R^\mu B_{\mu R} + \frac{M^2_C}{2}C^\mu C_\mu,
\end{equation}
where the masses are given by
\begin{eqnarray}
M_{B_{L}} &=& 2 g_{L} V_{33}, \\
M_{B_{R}} &=& 2 g_R V_{33}, \\
M_C &=& \sqrt{2(g_L^2 + g_R^2)} V_{33}\,.
\end{eqnarray}
Here $V_{33}$ is the vacuum expectation value of $S_3$. Additionally, the massless gauge boson corresponding to the 
resulting $U(1)_{B-L}$ symmetry is given by the linear combination
\begin{equation}
Z^\mu_{BL} = c_{\theta_3} C_L^\mu + s_{\theta_3} C_R^\mu \quad \text{with} \quad \text{tan}{\theta_3} = \frac{g_L^2}{g_R^2} \,,
\end{equation}
where the $B-L$ gauge coupling is given by
\begin{equation}
g_{BL} = \frac{\sqrt{3}}{2}\frac{g_L g_R}{\sqrt{g_L^2 + g_R^2}}\,.
\end{equation}
See Refs.~\cite{333-1,333-2,333-3,333-4,333-5,333-6} for several studies 
of theories based on trinification.
%
\section{433 at Colliders}
%
We have discussed above that the gauge theory based on $SU(4)_C \otimes SU(3)_L \otimes SU(3)_R$ 
predicts the existence of leptophobic gauge bosons and of stable colored or electric charged fermionic fields.
See Ref.~\cite{CMS} for the current experimental bounds on the leptophobic gauge boson assuming that it decays only into SM quarks, 
Ref.~\cite{vector-like} for the searches of vector-like fermions and Ref.~\cite{LLCP} for the signatures of stable charged particles. 

In Fig.~1 we show the total decay width of the leptophobic gauge boson $Z_B$ as a function of the mass, 
using as input values $m_D=1$ TeV and $m_{\Psi,\eta}=600$ GeV. We find that the total 
decay width can be large.  
\begin{figure}[ht]
	\centering
		\includegraphics[width=0.48\textwidth]{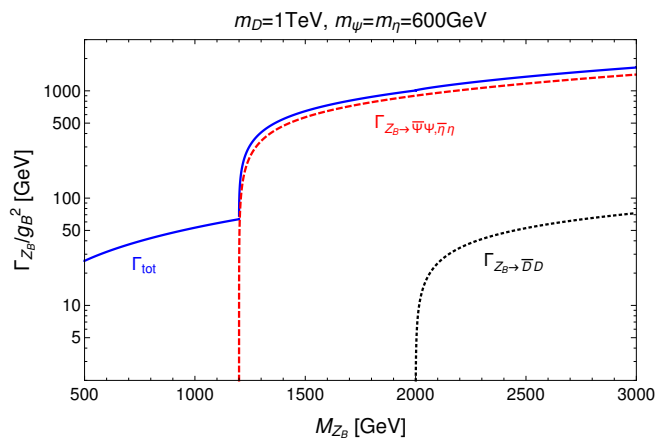}
		\caption{The decay width of the leptophobic gauge boson $Z_B$.}
\end{figure}

In Fig.~2 we show the branching ratios of the gauge boson $Z_B$ taking into account the decays into all SM quarks, 
into two $D$-quarks and into fractional charged fields. Using the same input parameters as in Fig. 1 we show 
that the decays into fractional charged fermions have the largest branching due to the fact that 
they have larger baryon number. This is an interesting result because one could expect that $Z_B$ could decay mainly into quarks. 
Since the branching ratio into fractional charged fermions is very large all the existent experimental bounds are modified. 
\begin{figure}[ht]
	\centering
		\includegraphics[width=0.48\textwidth]{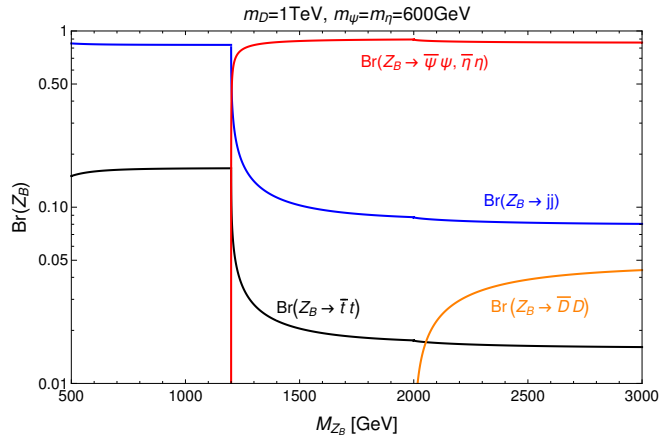}
		\caption{The branching ratios of the leptophobic gauge boson $Z_B$.}
\end{figure}
There are also other exotic signatures in this model:

\begin{itemize}

\item In the case when one of the new colored fields, $X_\mu$ or $\Phi_i$, is stable we can have the pair production through the QCD interactions. 
Then, one can have the formation of R-hadrons which leave very striking signals in the LHC detectors. See Ref.~\cite{LLCP} for the current bounds coming from these searches at the LHC. 

\item In the case when the stable particle is a fractional charged particle one predicts the existence of charged tracks.  
See Ref.~\cite{LLCP} for related searches at the LHC.

\end{itemize} 
In this model the fractional charged particles can be produced through the SM gauge bosons and the new leptophobic gauge boson $Z_B$, i.e. $pp \to \gamma^*, Z^*, Z_B^* \to \overline{\Psi} \Psi, \overline{\eta} \eta$, giving rise to two charged tracks. The existence of a new resonance allows us to have a larger cross section and thus the experimental bounds in Ref.~\cite{LLCP} will be stronger. 
These exotic signatures are crucial to identify this theory at current or future collider experiments.  
%
\section{444 Unification}
%
The gauge group $G_{433}$ is the symmetry group with the lowest rank which can embrace the Standard Model symmetries and gauged baryon number into a non-Abelian group. However, the symmetry group
\begin{equation}
G_{444} = SU(4)_C \otimes SU(4)_L \otimes SU(4)_R \,,
\end{equation}
can also accomodate for the Standard Model symmetries and gauged baryon number, and additionally can have a $Z_3$ symmetry identifying $SU(4)_C$, $SU(4)_L$ and $SU(4)_R$. The Standard Model fermions are embedded into 
\begin{equation}
Q \sim (4, \bar{4}, 1) \,, \quad Q^c \sim (\bar{4}, 1, 4) \,, \text{and} \quad L \sim (1, 4, \bar{4})  \,,
\end{equation}
where no new fields have to be added for anomaly cancellation. The exotic fields $\Psi^c$ and $\eta$ are embedded into $L$. Baryon number is again enclosed in $SU(4)_C$ which is broken by $(15, 4, \bar{4})_S$. Baryon number and the residual symmetries of $SU(4)_L \otimes SU(4)_R$ are broken by $(4, 1, 1)_S$, $(1, 4, 1)_S$ and $(1, 1, 4)_S$. To give the Standard Model quarks the measured masses we further have to add at least two scalars $(1, 4, \bar{4})_S$. However, the mass separation of the new heavy quarks and the light Standard Model quarks leads to large Yukawa couplings. Renomalizable lepton masses can be generated by a scalar bi-sextet $(1, 6, 6)_S$ or/and scalar bi-tenplet $(1, \bar{10}, 10)_S$. To generate the Standard Model lepton masses we find in both scenarios the need of large leptonic Yukawa couplings. This theory is very appealing but unfortunately one needs large Yukawa couplings, 
i.e. one can have problems with Landau poles,  to generate fermion masses in a consistent way. We will investigate this theory in great detail in a future publication.   
It is important to mention that in this context one can have the following symmetry breaking path:
$$4_C 4_L 4_R \to 4_C 3_L 3_R \to 3_C 2_L 1_Y 1_B \to \rm{SM}.$$
%
\section{Summary}
%
We have proposed the first UV completion of the models proposed in 
Refs.~\cite{B1,B2,B3,B4} where the baryon number is a local symmetry spontaneously broken at the low scale.
This theory is important to understand the possibility to have unification of gauge interactions at the low scale.
In this context the color and baryon number is unified in $SU(4)_C$ and 
the symmetry of the theory is $SU(4)_C \otimes SU(3)_L \otimes SU(3)_R$, which is broken to $3_C 3_L 3_R 1_B$ 
once $\Sigma \sim (15,1,1)$ acquires a vacuum expectation value. Now, the triplets 
$\Phi_i \sim (1,3,\bar{3})$ can break $3_C 3_L 3_R 1_B$ to the Standard Model plus the 
extra $U(1)_B$. However, in order to generate masses for the leptophobic gauge boson 
$Z_B$  and the extra fermions one needs to include in the Higgs sector 
a field in the fundamental of $SU(4)_C$. Thus, we find the first full non-Abelian 
gauge theory where one has spontaneous baryon number violation.

We have discussed the properties of all extra fields present in the theory and show 
that the theory predicts always a stable colored or with fractional electric charged particle.
This is clearly a problem for cosmology. In order to avoid this issue we have assumed that 
the reheating temperature is below the mass of the exotic fields. These fields can give rise 
to exotic signatures at the LHC since one can have charged tracks or formation 
of R-hadrons in case the stable fields are colored. We have discussed the properties 
of the leptophobic gauge bosons which are crucial to understand the testability of these theories. 
These theories can be the key for the realization of the unification of gauge interactions at the low scale 
since the proton decay is highly suppressed. Finally, we have discussed the embedding in a simple theory 
based on the $SU(4)_C \otimes SU(4)_L \otimes SU(4)_R$ gauge symmetry.

{\textit{Acknowlegments}: P.F.P. thanks M.B.Wise for discussions.}



\end{document}